\documentclass[twocolumn,showpacs,preprintnumbers,amsmath,amssymb,pra]{revtex4}

\usepackage{graphicx}
\usepackage{dcolumn}
\usepackage{bm}
\usepackage{amsmath}
\usepackage{float}
\usepackage{mathrsfs}
\usepackage{color}
\usepackage{color}
\newcommand{\be}{\begin{eqnarray}}
\newcommand{\ee}{\end{eqnarray}}

\begin{document}

\title{Comprehensive Study of Properties of a Endohedrally Confined Ca Atom using Relativistic Many-body Methods}

\author{S. Bharti$^{1}$, L. Sharma$^{1}$, B. K. Sahoo$^{2}$, P. Malkar$^{1}$ and R. Srivastava$^{1}$ }

\affiliation{
$^1$Department of Physics, Indian Institute of Technology Roorkee, Roorkee 247667, India\\
$^2$Atomic, Molecular and Optical Physics Division, Physical Research Laboratory, Navrangpura, Ahmedabad 380009, India}

\date{\today}

\begin{abstract}
We have carried out theoretical investigations of electron correlation effects on the atomic properties 
of the Ca atom trapped inside an attractive spherically symmetric potential well of an endohedral fullerene C$_{60}$ cluster. Relativistic 
coupled-cluster (RCC) theory has been employed to obtain electron correlation energy, ionization potential and dipole polarizability of this 
atom. We have also performed calculations using the Dirac-Hartree-Fock (DF), relativistic second-order many-body perturbation theory (RMBPT(2)
method) and relativistic random phase approximation (RRPA) to demonstrate propagation of the correlation effects in these properties. Our 
results are compared with the reported calculations employing multi-configuration Hartree-Fock (MCHF) method in Phys. Rev. A {\bf 87}, 013409 
(2016). We found trends in correlation energy with respect to the potential depth are same, but magnitudes are very large in the relativistic 
calculations. We have also determined the differential and total cross-sections for elastic scattering of electrons from 
the free and confined Ca atoms using the electronic charge densities from the Dirac-Hartree core-potential (DFCP) and 
RCC methods to demonstrate role of potential depth in these 
properties.
\end{abstract}


\maketitle

\section{Introduction}

Studies related to confined atoms are important for a broad range of possible applications. The spatial entrapment of atoms or molecules is
possible in nature as well as artificially;  e.g., impurity 
atoms in mesoscopic scale semiconductor artificial structures, molecular zeolite sieves, fullerenes, quantum devices, etc.  \cite{imp-1,imp-2}. 
The confinement of atoms lead to distinct and interesting changes in their electronic structures and other properties. Endohedral entrapment of 
an atom inside fullerene cage can be synthesized in laboratory owing to rapid developments in the modern technologies. These endohedral 
fullerenes unveil interesting physical and chemical properties that can be significantly different from the encapsulated atom or the fullerene 
itself. Although studies of confined endohedral atoms in any fullerene are interesting, but confined atoms particularly in the C$_{60}$ 
fullerene has drawn more attention these days. This is because C$_{60}$ can be assumed as a spherical ball with the confined atom at the center.
Therefore, the energy levels of a confined atom can still be described using a spherical potential as for a free atom by suitably choosing an 
effective potential.

Using R-matrix method in the non-relativistic theory, Haso\u glu {\it et al.} \cite{STM-1} had calculated cross-sections for the photoionization
process of a confined Ca atom and reported the influence of potential well depth on near threshold photoionization cross-sections. Following 
this work, Kumar et al \cite{HV-1} had investigated non-dipole effects in photoionization of outer 4$s$ shell of the entrapped Ca atom in a 
spherical attractive well potential in the relativistic random phase approximation (RRPA). Recently, Haso\u glu {\it et al.} \cite{Zuo} have 
investigated electron correlation effects of endohedral confinement on the energy levels of Be, Mg, and Ca atoms in the non-relativistic 
framework using multi-configuration Hartree-Fock (MCHF) method. Haso\u glu {\it et al.} have argued in their work about the possibility of 
confinement of Ca atom at the center of the C$_{60}$ shell, giving rise to a stable equilibrium. It is, therefore, imperative to understand 
energy levels of such an environment on the structure of a captured Ca atom more accurately by employing a powerful relativistic 
many-body method before carrying out experiment.  

It is very challenging to get accurate results for the atomic properties of the Ca atom even when it is isolated from any external mediation 
owing to strong correlations between its two valence $s$ electrons. For example, it has been found earlier that the lower-order relativistic 
perturbative theories underestimate the electric dipole polarizability ($\alpha_0$) value of the Ca atom while RRPA overestimates it 
compared to the experimental value \cite{yashpal1}. As mentioned above, only a few less powerful many-body methods are employed to account for 
electron correlation effects in the confined Ca atom. To guide future experiment on the confined Ca atom, it is necessary to study its
spectroscopic properties by applying a more accurate theory. In this work, we have calculated many properties such as electron correlation 
energy, ionization potential (IP) and $\alpha_0$ of the of the confined Ca atom in the endohedral fullerene C$_{60}$ cluster 
and show their variation with the attractive potential depth. These calculations are performed within the framework of the Dirac-Hartree-Fock 
(DF) method, relativistic second-order many-body perturbation theory (RMBPT(2) method), RRPA and relativistic coupled-cluster (RCC) theory.
We employ the lower-order methods like DF and RMBPT(2) methods in order to find out role of electron correlations in the above properties 
by comparing results from these methods. The RCC theory is an all-order perturbative method 
that obeys proper scaling with the number of electrons even when it is approximated in contrast to the truncated configuration interaction 
(CI) method or its variants like the MCHF method \cite{szabo,bartlett,bishop}. It also includes core-polarization effects to all-order as 
in RRPA along with other physical effects. To ascertain accuracies of the results, we first perform calculations of the aforementioned properties 
of the free Ca atom and compare them with their corresponding experimental values. We also compare our calculations of self-consistent field 
(SCF) energy and energies of the 4$s$, 3$p_{3/2}$, 3$p_{1/2}$ and 3$s$ orbitals at various depths of the confining annular potential well with 
the earlier reported results by Kumar {\it et al.} using the DF method \cite{HV-1}. 

Because of experimental feasibility to prepare confined atoms, the interest in electron elastic collision with entrapped atoms is increasing 
now-a-days \cite{Dolmatov-1,Dolmatov-2,Fassen1,Fassen2}. However, there have been few theoretical studies reported in the context of elastic 
scattering of electrons from confined atom A@C$_{60}$. Dolmatov et al. have studied elastic collision of low energy electrons with A@C$_{60}$
and demonstrated the role of polarizability of an atom $+$ C$_{60}$ \cite{Dolmatov-1} and that of atoms alone \cite{Dolmatov-2}. In fact, 
Dolmatov {\it et al.} have given a full account of the previous related works in Refs. \cite{Dolmatov-1,Dolmatov-2} on the same. However, 
these studies were carried out using non-relativistic Hartree-Fock theory and at low projectile electron energies. Here, we have performed 
calculations of the elastic collision cross-sections for the intermediate projectile electrons scattered from the endohedrally confined Ca atoms 
in the relativistic theory framework. Furthermore, we have used electron charge densities from the DF method, DF method with core-polarization 
(CP) potential (DFCP method) and RCC theory to show the differences in the results due to various approximations in the many-body methods. 
It to be noted that RCC theory includes CP potential to all-orders along with electron correlations due to non-CP type of interactions. In all 
these approaches, we take into account the quantum mechanical exchanges between the projectile and atomic electrons as well as polarizability of
the atom as a function of potential depth. Since we focus only on the effect of confinement of Ca on electron scattering process, we have 
neglected the impact of polarizability of C$_{60}$ by the incident electron in these calculations. The results are reported for differential 
cross-sections (DCSs) and integrated cross-sections (ICSs). We compare our results for the free Ca atom with the available experimental 
data \cite{Ca-expt} and other reported calculations using the DFCP method \cite{hasan} in order to gauge accuracies of the results for the 
confined atom.  

\section{Methods for Calculations}
\subsection{Atomic properties}

In the DF method, we obtain the mean-field wave function $|\Phi_0 \rangle$ using the $V^N$ potential of the $[3p^{6}4s^2]$ configuration of 
the Ca atom. All the DF and DFCP results are obtained using this wave function and $\alpha_0= 169\ ea_0^3$ from experiment \cite{exp-pol}. The 
exact wave function in the RCC theory is expressed using $|\Phi_0 \rangle$ as the reference determinant as \cite{cizek}
\begin{equation} \label{exp}
| \Psi_0 \rangle = e^{\hat{T}} | \Phi_0 \rangle ,
\end{equation}
where $\hat T$ is known as the RCC excitation operator given by
\begin{eqnarray} \label{Texp}
\hat T  &=&  \sum\limits_{k=1}^N  \hat T_k = \sum\limits_{\stackrel{a_1 < a_2 \dots < a_k}{i_1 < i_2 \dots < i_k}} t^{a_1 a_2 \dots a_k}_{i_1 i_2 \dots i_k}   a^+_1 i^-_1 a^+_2 i^-_2 \dots a^+_k i^-_k  \ \ \ \
\end{eqnarray}
with $+$ and $-$ superscripts on the second quantization operators represent for creation and annihilation of electrons in the virtual (denoted by $a$) and occupied (denoted by $i$) orbitals,
respectively, and $t$ are the amplitudes in the excitation process in an $N$ electron system. The RCC approaches considering up to $T_N$ operators with $N = 2, 3, 4, \dots$, known as the RCC 
singles and doubles (RCCSD), RCC singles, doubles, and triples (RCCSDT), RCC singles, doubles, triples, and quadruples (RCCSDTQ), etc. methods constitute a hierarchy, which converges to the exact 
solution of the wave function in the given one-particle basis set.

 The amplitudes $t$ of the RCC operators are obtained by projecting bra determinants $\langle \Phi^{a_1 a_2 \dots a_k}_{i_1 i_2 \dots i_k} | e^{-\hat{T}}=$ $\langle \Phi_0^N | a^+_1 i^-_1 a^+_2 i^-_2 \dots  a^+_k i^-_k e^{-\hat{T}} $ from the
left of the Schr\"odinger equation $\hat H |\Psi_0 \rangle = E_0 |\Psi_0 \rangle$, with the ground state energy $E_0$, as
\begin{eqnarray}\label{CCeq}
\langle \Phi^{a_1 a_2 \dots a_k}_{i_1 i_2 \dots i_k} \vert \overline{H} \vert \Phi_0 \rangle &=& E_0 \delta_{k,0}, \;\;\;\; (k=1, \dots N) ,
\end{eqnarray}
where $ \overline{H}=e^{-\hat{T}} \hat H e^{\hat{T}}=(\hat H e^{\hat{T}})_c$ for the subscript $c$ means connected terms between the atomic 
Hamiltonian $\hat H$ with the $\hat T$ operators are only retained.

We have considered the Dirac-Coulomb (DC) Hamiltonian in our calculation for the free atom, which in atomic units (a.u.) is given by   
\begin{eqnarray}
H &=& \sum_{i=1}^{N} \left [ c\mbox{\boldmath$\alpha$}_i\cdot \textbf{p}_i+(\beta_i -1)c^2 + V_{n}(r_{i}) \right] 
   +  \frac{1}{2} \sum_{i,j} \frac{1}{r_{ij}}  \nonumber \\
  \label{NSDeq}
\end{eqnarray}
where $\mbox{\boldmath$\alpha$}$ and $\beta$ are the Dirac matrices, $c$ is the speed of light, the nuclear potential $V_{n}(r)$ is determined 
using the Fermi-charge distribution, and $r_{ij}$ is the radial distance between the electrons located at $r_i$ and $r_j$. To obtain wave
function of the endohedral confined atom, we introduce a short range spherical model potential $U$ along with the DC Hamiltonian defining as 
\be 
U(r) =
\left\{\begin{array}{ll}
 -U_0, & \mbox{for $r_0 \le r \le r_0 + \Delta $}\\
0, & \mbox{otherwise}
\end{array}\right.
\label{Ur}
\ee
where $r_0$ is the inner radius, $\Delta$ is the thickness and $U_0$ is the depth of the potential well.

The IP of an electron of orbital $a$ in the atom is estimated in the RCC theory by expressing \cite{nandy1}
\begin{eqnarray}
|\Psi_a \rangle = e^T (1+ R_a) |\Phi_a \rangle,
\end{eqnarray}
where the reference state is constructed as $|\Phi_a \rangle = a_a |\Phi_0 \rangle$ with $a_a$ is the corresponding annihilation operator 
for the electron in orbital $a$ and $R_a$ is another RCC operator introduced to take care of the extra correlation effects that was 
included through the detached electron. The IP ($E_a$) and amplitude solving equations for the $R_a$ wave operators 
are given by
\begin{eqnarray}
\langle \Phi_a^*| [(H e^T)-E_a] \{1+R_a\}|\Phi_a\rangle &=& \delta_{a,*}
\end{eqnarray}
where $|\Phi_a^* \rangle$ designates excited configuration determinants from $|\Phi_a\rangle$ for the $R_a$ amplitude determination else 
it corresponds to $|\Phi_a\rangle$ to estimate IP of the electron from orbital $a$. This equation contains non-linear terms and also both 
the energy and amplitude determining equations are inter-dependent. These solutions are obtained adopting the self-consistent procedure.

The excited states ($|\Psi_K(J, \pi) \rangle$) of the Ca atom with angular momentum $J$ and parity $\pi$ is obtained by operating 
excitation operators $\Omega_K$ on $|\Psi_0 \rangle$ as \cite{nandy2} 
\begin{eqnarray}
 |\Psi_K(J, \pi) \rangle = \Omega_K(J, \pi) |\Psi_0\rangle ,
\end{eqnarray}
where $K$ corresponds to level of excitations. The eigenvalue ($E_L$) and eigenfunction for the $L^{th}$ excited state are obtained 
by diagonalizing the above equation as
{\small{
\begin{eqnarray}
\langle \Phi_L (J, \pi) | (H e^T) \Omega_K (J, \pi) |\Phi_0 \rangle = E_L \langle \Phi_L (J, \pi) | \Omega_L (J, \pi) |\Phi_0 \rangle .
\end{eqnarray}}}
In this expression, $|\Phi_L\rangle$ has a definite value of $J$ and $\pi$ and solutions are obtained in the iterative procedure using the 
Davidson's diagonalization algorithm for the non-symmetric matrix \cite{hirao}.

We evaluate $\alpha_0$ by expressing the RCC wave function as \cite{bijaya1}
\begin{eqnarray}
 |\tilde{\Psi}_0 \rangle =e^{\hat{T}^{(0)}+ |\vec{\mathcal E}| \hat{T}^{(1)}} |\Phi_0 \rangle,
\end{eqnarray}
where $\hat{T}^{(0)}$ represents the RCC operator that accounts for electron correlation effects due to the electromagnetic interactions 
only and $\hat{T}^{(1)}$ takes care of correlation effects due to both the electromagnetic interaction and the $D$ operator, respectively, 
to all-orders. In the perturbative expansion, this corresponds to
\begin{eqnarray}
 |\Psi_0^{(0)} \rangle = e^{\hat{T}^{(0)}} |\Phi_0 \rangle
\ \ \ \ 
\text{and}
\ \ \ \
 |\Psi_0^{(1)} \rangle = e^{\hat{T}^{(0)}} \hat{T}^{(1)} |\Phi_0 \rangle.
 \label{eqt1}
\end{eqnarray}
Both $|\Psi_0^{(0)} \rangle$ and $|\Psi_0^{(1)} \rangle$ can be determined by obtaining amplitudes of the $\hat{T}^{(0)}$ and $\hat{T}^{(1)}$ 
RCC operators. The amplitude determining equation for $\hat{T}^{(0)}$ is same as Eq. (\ref{CCeq}) for the DC Hamiltonian. The $\hat{T}^{(1)}$ 
amplitude determining equation is given by
\begin{eqnarray}
 \langle \Phi^{a_1 a_2 \dots a_k}_{i_1 i_2 \dots i_k} \vert \overline{H}^{DC} \hat{T}^{(1)}+ \overline{D} \vert \Phi_0 \rangle =0 \label{eq37} .
\end{eqnarray}
It to be noted that for solving the amplitudes of $\hat{T}^{(0)}$, the projected $\langle \Phi^{a_1 a_2 \dots a_k}_{i_1 i_2 \dots i_k} \vert$ 
determinants have to be even parity whereas they are the odd-parity for evaluating the $\hat{T}^{(1)}$ amplitudes. We have considered 
the RCCSD method approximation in the RCC theory.

After obtaining these solutions, we evaluate $\alpha_0$ as \cite{yashpal1,yashpal2}
\begin{eqnarray}
\alpha_0  &=& 2 \langle\Phi_0 | e^{T^{(0)\dagger}} D e^{T^{(0)}} T^{(1)} | \Phi_0 \rangle_{fc} ,
\label{eqccx}
\end{eqnarray}
where $fc$ stands for the fully-contracted terms. The above expression contains a non-terminating series $e^{T^{\dagger (0)}} D e^{T^{(0)}}$. This is computed self-consistently as discussed in 
our earlier works \cite{yashpal2,bijaya2}. We also perform calculation of $\alpha_0$ using the RMBPT(2) and RRPA methods adopting the procedures 
described in Refs. \cite{yashpal1,yashpal2,yashpal3}. From the differences between the results obtained by the RRPA and RCCSD method, we can find 
out contributions due to the non-core-polarization correlations to all-orders.

\begin{figure*}[t]
\begin{center}
\includegraphics[trim = 0cm 5cm 5cm 0cm,scale=0.7]{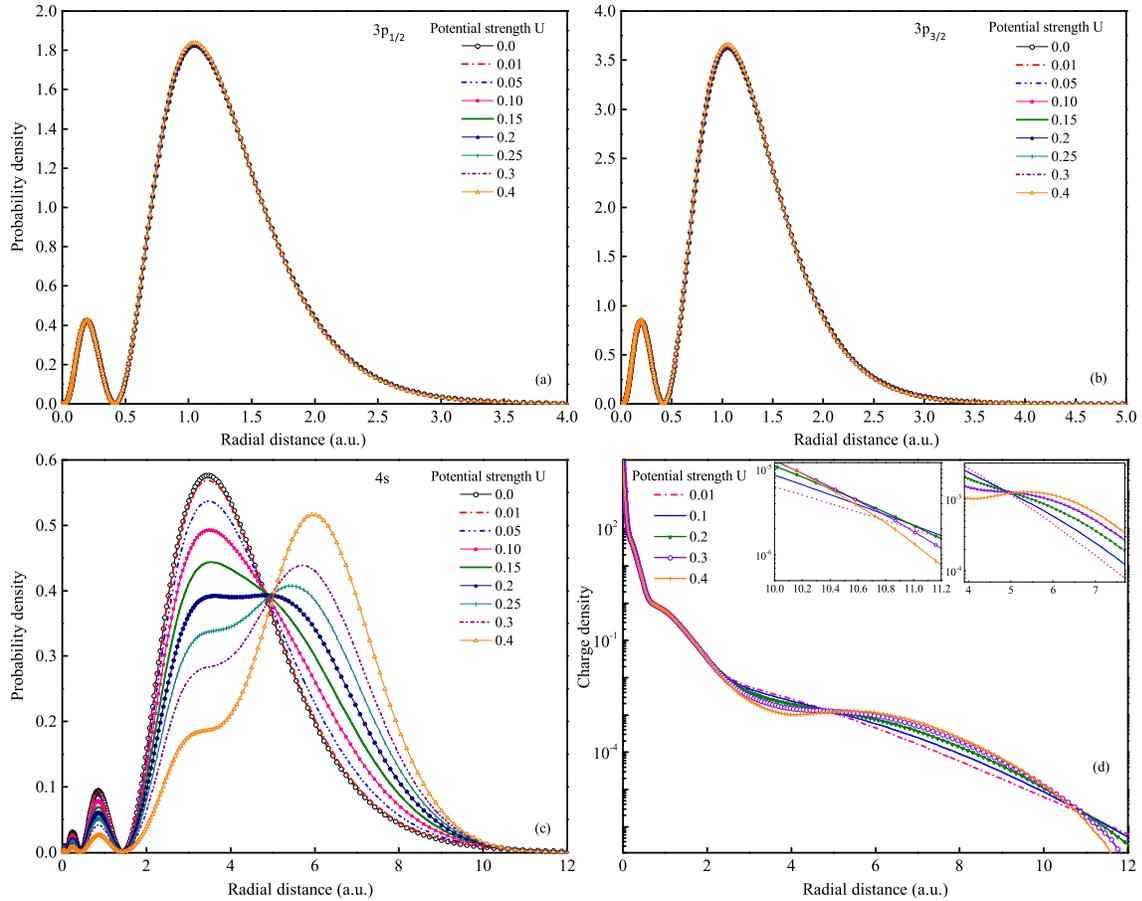}
\end{center}
\caption{(Color online) Probability densities of the 3$p_{1/2}$, 3$p_{3/2}$ and 4$s$ orbitals are shown in (a), (b) and (c), respectively,
and (d) shows charge density ($\rho(r)$)  as function of the radial distance ($r$) and  the confining potential $U$ (in a.u.). 
The insets in (d) magnify the crossing of curves around 5 a.u. and large radial distances.}
\label{wavefunctions}
\end{figure*}
\subsection{Cross-section calculations}
\label{Model-Potential-theory}

If the interaction potential $V(r)$ of the projectile electron with the target is approximated to be 
spherically symmetric then the method of partial wave expansion can be employed and the entire scattering process
can be described by the direct and spin-flip scattering amplitudes as 
\be
\label{direct}
f(k,\theta) &=& \frac{1}{2 \iota k} \sum_{l=0} ^\infty \left (
(l+1) (e^{2\iota \delta^+} -1) \right. \nonumber \\
&& + \left. l (e^{2\iota \delta^-} -1) \right) P_l(\cos\theta)
\ee 
and
\be
\label{spin-flip}
g(k,\theta) = \frac{1}{2 \iota k} \sum_{l=0} ^\infty \left (
 e^{2\iota \delta^-} -e^{2\iota \delta^+} \right)
 P_l^1(\cos\theta),
\ee  
respectively. Here $k$ is the relativistic wave number, $\delta^{\pm}$ are the scattering phase shifts with  
$+$ ($-$) sign refers to up (down) spin of the projectile electron, $\theta$ is the scattering angle, and $P_l(\cos\theta)$ and 
$P_l^1(\cos\theta)$ are the Legendre polynomials and associated Legendre functions, respectively. The phase shifts are calculated by 
solving the Dirac radial equation in the field of $V(r)$ and analyzing the large-$r$ behavior of the Dirac spherical waves. 

The DCS per unit solid angle for spin unpolarized electrons are calculated by
\be
\label{dcs}
\frac{d\sigma}{d\Omega} = |f(k,\theta)|^2 + |g(k,\theta)|^2 .
\ee
We obtain the ICS by integrating the DCS over the solid angle. For the computational simplicity,
we express the interaction potential as $V(r) = V_{st}(r) + V_{ex}(r)- \iota V_{ab}(r)$, where $V_{st}$, $V_{ex}(r)$ and $V_{ab}$ are
known as the static, exchange and absorption potentials, respectively. The static potential at a general point $r$ is evaluated using 
the following expression
\be 
\label{vstatic}
V_{st}(r) = -\frac{Z}{r} -\left(\int_0^r \rho(r') 4\pi r'^2 dr' 
                        + \int_r^\infty \rho(r') 4\pi r' dr'                                            
                        \right).
\ee
Here $Z$ is the nuclear charge of the atom and $\rho(r)$ is the charge densities of the confined Ca atom. We determine $\rho(r)$ first 
in the mean-field procedure using the DF method and then, we take into account the electron screening effects through the RCC method by 
including the electron correlation effects. After evaluating $V_{st}(r)$, the exchange potential is calculated using 
the charge density and energy dependent form suggested
by Furness and McCarthy \cite{Ve_FM}. In the elastic scattering process, there is still finite probability of occurrence of 
inelastic processes which we take into 
account through the absorption potential $V_{ab}(r)$. We define this potential in the local density approximation (LDA) as suggested by 
Salvat \cite{salvat}. 

As some of the previous studies were carried out using the DFCP method (e.g. see Refs. \cite{hasan,ls-1,ls-2}), we also employ the DFCP method in our 
calculations to demonstrate differences between the results from this method and the RCC theory by introducing the CP 
potential ($V_{cp}(r)$) in the interaction potential as $V(r) = V_{st}(r)+V_{ex}(r)+V_{cp}(r)-\iota V_{ab}(r)$. Here $V_{st}(r)$ and 
$V_{ex}(r)$ are determined using $\rho(r)$ from the DF method. This approximation for $ V(r) $ may suit better for slow incident electrons as 
they can polarize the charge cloud of atom more causing interaction of the induced dipole moment of the atom with the projectile electron. 
The CP potential can be divided into two parts as short-range ($V_{sr}$) and long range ($V_{lr}$)  as discussed by Salvat \cite{salvat}. 
Here, we define $V_{sr}(r)$ in LDA  \cite{lda} considering the parameters given by Pedrew
and Zunger \cite{cp_sr}. Buckingham potential \cite{Buckingham} is employed to describe $V_{lr}(r)$ which 
depends upon the static dipole polarizability $ \alpha_{0} $. It is evident now that to define interaction potential $V(r)$ reliably 
it is necessary to obtain $\rho(r)$, $\alpha_0$ and also excitation energies (which are required to determine $V_{ab}(r)$) as accurately
as possible. Moreover, their dependencies on the well depth of the endohedral confinement need to be investigated precisely in order 
to study elastic scattering from the confined atom.

\section{Results and Discussion}\label{RnD}

\subsection{Atomic properties}

In the present work, we model the endohedral confinement potential in Eq.~(\ref{Ur}) by using the values of $r_0$ and $\Delta$ 
to be 5.8 a.u. and 1.89 a.u., respectively, as suggested by Xu et al \cite{Xu}. As mentioned earlier, Haso\u glu {\it et al.} 
\cite{STM-1} had investigated spectroscopic properties of the confined Ca atom using the MCHF method. Here, we intend to show the  
differences in some of the results due to our relativistic methods compared to the MCHF calculations. We use Gaussian type orbitals 
(GTOs) in the DF method to obtain the single particle orbital wave functions and energies. The $k^{th}$ GTO in an atomic orbital expansion is 
defined as 
\be
f_k(r) = e^{-\eta_k r^2},
\ee
where $\eta_k$ is an arbitrary parameter chosen suitably to produce the orbitals accurately. Using the even tempering condition as 
$\eta_k = \eta_0 \zeta^k$, we consider $\eta_0=0.00715$ and $\zeta=1.92$ to define GTOs in our calculations. From these calculations, we 
observe that increasing the potential depth does not affect much the inner shell orbitals wave functions. However, for outer shell 4$s$ the effect
of change in the value of $U$ is quite significant; more pronounced in the vicinity of the radial distances $r_0 -\Delta$ and 
$r_0 +\Delta$. To demonstrate this, we have plotted variation of probability densities of the inner 3$p_{1/2}$ and 3$p_{3/2}$ and outer 4$s$ shells
in Figs. \ref{wavefunctions}(a)-(c), respectively, from the DF method. As seen, the peak of the probability density curve shifts for the $4s$ 
orbital towards larger $r$ as the well deepens. All the curves cross one another nearly at inner radius $r_0$ of the potential well. Our 
results are in very good agreement with the numerical calculations by Kumar et. al. \cite{HV-1}. Figure \ref{wavefunctions}(d) shows the 
electronic charge density  $\rho(r)$ as a function of $r$ at various depths. This shows the probability density of the 4$s$ orbital, $\rho(r)$, 
also changes dramatically close to the inner-radius of the potential well. In the inset of Fig. \ref{wavefunctions}(d), we show how all the 
curves cross approximately around $r = 5$ a.u.. This shows with increasing $U$ value, $\rho(r)$ first decreases slowly up to $r = 10$ a.u. 
and falls off faster beyond 10 a.u.. After discussing wave functions, we show variation of the single particle orbital energies 
with the potential depth. In this case also we find that the trends match well with the results of Kumar et. al. \cite{HV-1} as quoted 
in Table \ref{tab1}. This implies that we have correctly implemented the confined potential in our calculations.

\begin{table}
\caption{Comparison of single particle orbital energies from our analytical calculations using GTOs (given as I) and numerical values reported 
by Kumar et al \cite{HV-1} (given as II) for different confined potential strength U (in a.u.).}
\begin{tabular}{c c c c c c c c c}\hline \hline
U & \multicolumn{2}{c}{4$s$}     & \multicolumn{2}{c}{$3p_{1/2}$} & \multicolumn{2}{c}{$3p_{3/2}$}& \multicolumn{2}{c}{$3s$} \\
         & I & II & I & II & I & II & I & II \\ \hline
         & & \\
0.000	& 5.34	& 5.34		&36.71	& 36.71	& 36.28	& 36.29	& 61.55	& 61.55 \\
0.005	& 5.37	& 5.37		&36.75	& 36.75	& 36.33	& 36.34	& 61.59	& 61.59 \\
0.010	& 5.40	& 5.40		&36.79	& 36.79	& 36.37	& 36.37	& 61.63	& 61.63 \\
0.050	& 5.65	& 5.65		&37.14	& 37.14	& 36.72	& 36.72	& 61.97	& 61.97 \\
0.090	& 5.93	& 5.93		&37.54	& 37.53	& 37.12	& 37.10	& 62.36	& 64.35 \\
0.130	& 6.24	& 6.24		&38.00	& 37.95	& 37.59	& 37.53	& 62.82	& 62.77 \\
0.140	& 6.35	& 6.33		&38.09	& 38.06	& 37.67	& 37.64	& 62.90	& 62.88 \\
0.170	& 6.62	& 6.60		&38.45	& 38.41	& 38.03	& 37.99	& 63.26	& 63.22 \\
0.200	& 6.91	& 6.88		&38.65	& 38.77	& 38.44	& 38.35	& 63.67	& 63.57 \\
0.200	& 7.02	& 6.98		&38.99	& 38.89	& 38.58	& 38.47	& 63.80	& 63.69 \\
0.220	& 7.13	& 7.09		&39.13	& 38.01	& 38.71	& 38.59	& 63.93	& 63.81 \\
0.250	& 7.47	& 7.41		&39.54	& 39.39	& 39.12	& 38.96	& 64.33	& 64.18 \\
0.290	& 7.96	& 7.87		&40.10	& 39.88	& 39.64	& 39.46	& 64.90	& 64.66 \\
0.310	& 8.23	& 8.11		&40.34	& 40.12	& 39.92	& 39.70	& 65.12	& 64.90 \\
0.330	& 8.50	& 8.36		&40.60	& 40.35	& 40.20	& 39.93  & 65.40& 65.13 \\\hline \hline
\end{tabular}
\label{tab1}
\end{table}

After testing single particle orbital energies and wave functions from our DF method, we now compare the electron self-consistent field (SCF) 
and correlation energies with the MCHF calculations in Table \ref{corr}. We find our SCF energy of the free Ca atom is about 3 a.u. larger than
the SCF energy reported using the MCHF method in Ref. \cite{Zuo}. We also perform numerical calculation of the SCF energy using the GRASP2K 
package \cite{grasp2k} to validate our result. There is a good agreement between both the calculations in the relativistic approach. This 
justifies that the difference between the SCF energies from our calculations and the MCHF results of Haso\u glu {\it et al.} \cite{Zuo} is due to the relativistic 
effects. Again, the correlation energies obtained using the RMBPT(2) and RCCSD methods for both the isolated and confined Ca atoms are found 
to be considerably different than the values given by the MCHF method. It suggests that the relativistic and correlation corrections 
are inefficiently included in the MCHF method. We also find that the correlation energies from the RMBPT(2) and RCCSD methods are very close 
to each other and they are decreasing with the increasing value of the potential depth. 

\begin{table}
\caption{Demonstration of trends of SCF and correlation energies (absolute values) as function of $U$. We also compare our SCF energy for 
the free Ca atom from the non-relativistic calculation reported by Haso\u glu et al \cite{Zuo} and using GRAS2K code \cite{grasp2k}. 
Similarly, magnitudes of the correlation energies are also compared between the calculations using the RMBPT(2), RCCSD and MCHF methods. All the quantities are 
given in a.u.}
\begin{tabular}{lccccc}\hline\hline
$U$ & \multicolumn{2}{c}{SCF Energy} & \multicolumn{3}{c}{Correlation Energy}\\\cline{2-6}
    &         This work & Others              & RMBPT(2)    & RCCSD & MCHF \cite{Zuo} \\\hline
    & & \\
 0.00       & 679.71  &  676.76 \cite{Zuo}      & 0.7302    &   0.7487	& 0.223\\
	    &  & 679.71 \cite{grasp2k} & & &\\
   0.01   &    676.79   & &  0.7297    &   0.7483		& 0.222\\
   0.05   &      679.72  &	&   0.7276   &     0.7465 	& 0.221\\
   0.10   &      679.74 &  &  0.7247   &    0.7442		& 0.219\\
    0.15   &     679.76   & &	 0.7218   &    0.7419	& 0.217\\
    0.20    &     679.79  &  &	 0.7189  &     0.7399	& 0.215\\
    0.25   &     679.82   &  &	0.7162    &   0.7379		& 0.213\\
    0.30   &      679.85  &   &	0.7138   &    0.7369		& 0.212\\
    0.35    &    679.89    &	&	0.7119    &   0.7379		& 0.211\\
    0.40     &    679.93  & &	  0.7107   &    0.7410 	& 0.210\\\hline \hline
\end{tabular}
\label{corr}
\end{table}

We now show the trends of IP of the confined Ca atom with well depth in Table \ref{ipeng} from the DF, RMBPT(2) and RCCSD methods. We have also given its
value for the free atom and compared with the experimental value \cite{exp-ip} in the same table. The RPMBT(2) result seems   
closer to the RCCSD result. We anticipate that inclusion of triple excitations in the RCC theory will improve the RCCSD result, but 
inclusion of these excitations demands for large computational resources. Furthermore, carrying out calculations including triple excitations 
for a range of well depths are too time taking. Thus, we show trends of the IP only using the RCCSD method approximation in the RCC theory 
framework. 

\begin{figure*}[t]
\begin{center}
\includegraphics[trim = 0cm 1cm 5cm 0cm,scale=0.8]{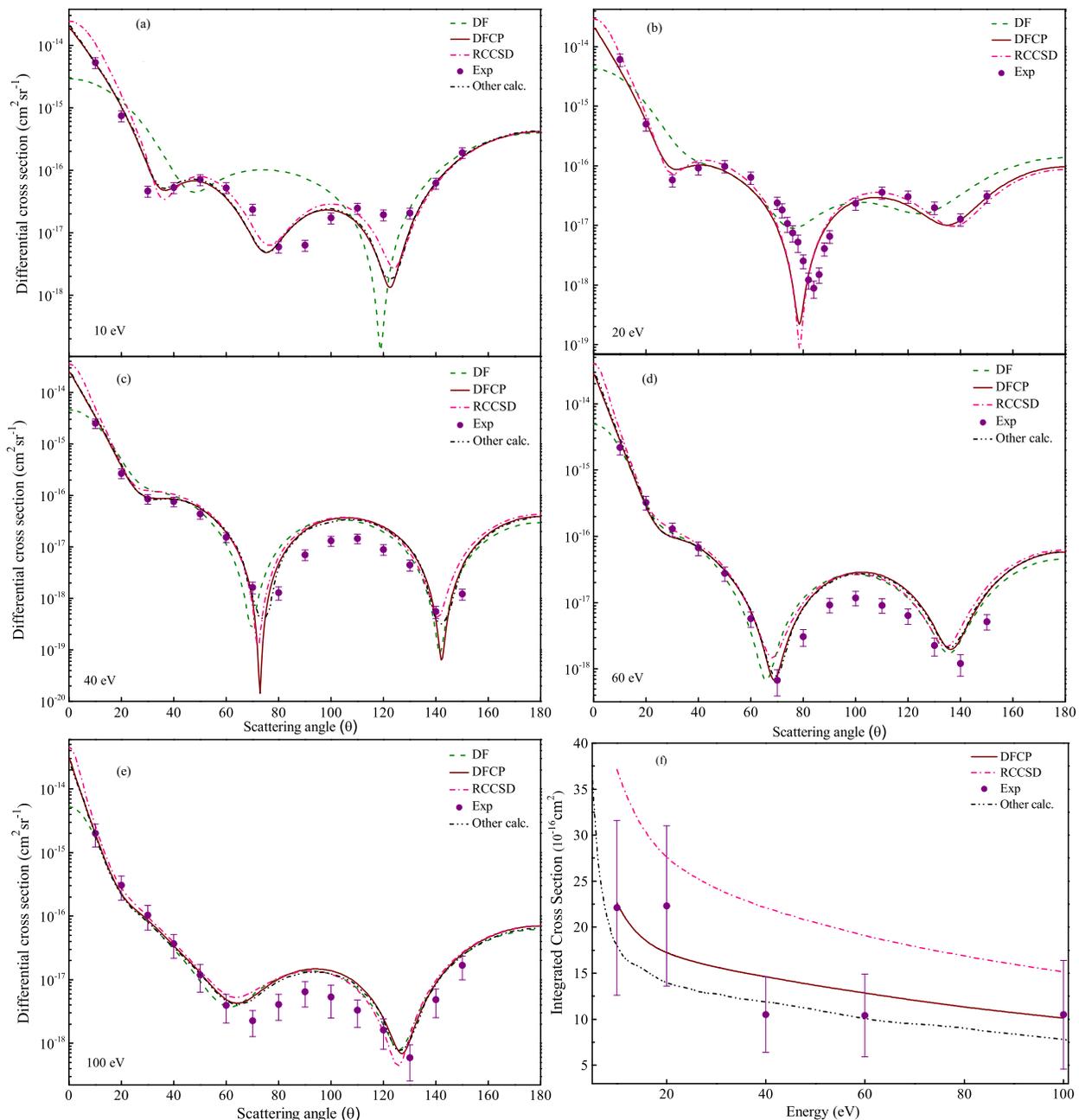}
\end{center}
\caption{(Color online) Comparison of our DCSs and ICSs from the DF, DFCP and RCCSD methods with the experimental value 
\cite{Ca-expt} and other available calculation \cite{hasan} for the elastic scattering of an electron from an isolated Ca atom.}
\label{free-ca}
\end{figure*}

\begin{table}
\caption{Trends of ionization potential (IP) in cm$^{-1}$ from the DF, RMBPT(2) and RCCSD methods as function of $U$ (in a.u.). We 
also compare our results for the free Ca atom with the experimental value.}
\begin{tabular}{c c c c c}\hline \hline
     $U$   &  DF       &    RMBPT(2)     &     RCCSD      &    Experiment \cite{exp-ip} \\\hline
 & & \\
    0.00 &    43085.69   &    49769.48    &     50138.51  &   49305.95 \\
    0.01 &  43544.77    &   50157.94     &    50520.24	&\\
  0.05   & 45528.67     &  51864.55      &   52180.95		&\\
   0.10  &   48356.92   &    54371.71    &     54654.27	&\\
   0.15  &   51588.43   &   57330.23     &    57645.79		&\\
   0.20  &   55227.19   &   60759.73     &    61198.35	&\\
   0.25  &   59264.59   &   64659.86     &   65301.45		&\\
   0.30  &    63682.83  &    69015.21    &     70031.23	&\\
  0.35   & 	68458.74     & 73801.28       &  75508.26		&\\
   0.40    &   73566.84    &  78987.77     &    81725.72	& \\\hline \hline
\end{tabular}
\label{ipeng}
\end{table}

\begin{table}[b]
\caption{Correlation trends of dipole polarizability of the free and confined Ca atoms using the DF, RMBPT(2), RRPA and RCCSD methods. 
Results are given in a.u.}
\begin{tabular}{c c c c c c}\hline \hline 
       $U$ &    Experiment \cite{exp-pol}   &   DF    &    RMBPT(2)   &  RRPA   &     RCCSD \\\hline
& & \\
    0.00    &       169$\pm$17  &    122.89   &    151.73  &    182.80 	&     157.03 \\
    0.01    &                  &       125.82  &     155.72  &    187.60  & 161.12\\
    0.05    &                  &       138.81  &     173.28  &    208.52  & 180.29\\
    0.10    &                &          158.06  &   198.98   &   238.60   &   209.61\\
   0.15   &                  &        180.85   &    228.86  &    272.71  &   245.47\\
    0.20   &                  &         207.13   &    262.56  &  310.07   &   288.69\\
    0.25  &                  &         236.54   &    299.30  &    349.55  &  Diverge\\
   0.30    &                  &         268.38   &    337.94  &  389.80   &  Diverge\\
   0.35   &                  &        301.61    &   377.09   &   429.48  &  Diverge\\
    0.40   &                  &          335.01  &     415.3  &   467.49  &   Diverge \\\hline \hline
    \end{tabular}
\label{dipp}
\end{table}

Determining $\alpha_0$ value of an atomic system accurately in the {\it ab initio} approach is very challenging. We have carried out 
calculations of $\alpha_0$ values for a variety of atomic systems precisely using our perturbative RCC method \cite{yashpal1,yashpal2,yashpal3}.
In our investigations, we had found RRPA can give $\alpha_0$ values of atomic systems with inert gas configurations very close to the 
calculations using the RCCSD method and experimental values \cite{bijaya3}. However, RRPA overestimates the $\alpha_0$ values in other 
atomic systems while the RCCSD method is able to give very accurate results \cite{yashpal3,bijaya3}. From the computational point of view, 
RRPA takes much less time than the RCCSD method to carry out these calculations. Since experimental value of $\alpha_0$ of free Ca atom is 
not available precisely, we perform calculation of this quantity using the DF, RMBPT(2), RRPA and RCCSD methods. These values are given in 
Table \ref{dipp} along with the experimental result. It shows the DF method gives a lower value of $\alpha_0$ in the free Ca atom, while its 
result increases gradually in the RMBPT(2) method and RRPA. However, the RCCSD method gives an intermediate value between the RMBPT(2) and 
RRPA calculations implying that there are strong cancellations between the core-polarization and other correlation effects in the RCCSD 
method. Our RCCSD result is also within the error bar of the experimental value \cite{exp-pol}. We find similar trends in the $\alpha_0$ 
values from the DF, RMBPT(2), RRPA and RCCSD methods for finite value of the potential depth of the confined Ca atom. We also see the $\alpha_0$ value of the confined 
Ca atom increases rapidly with increasing value of the potential depth and the perturbed amplitudes in the RCC method do not converge for 
relatively large magnitudes of the potential depth. In order to estimate this quantity accurately for large $U$ values, one can extrapolate 
our RCCSD results from the trends of the RMBPT(2) and RRPA calculations.  

\begin{figure*}
\begin{center}
{\includegraphics[trim = 0cm 5cm 5cm 0cm,scale=0.8]{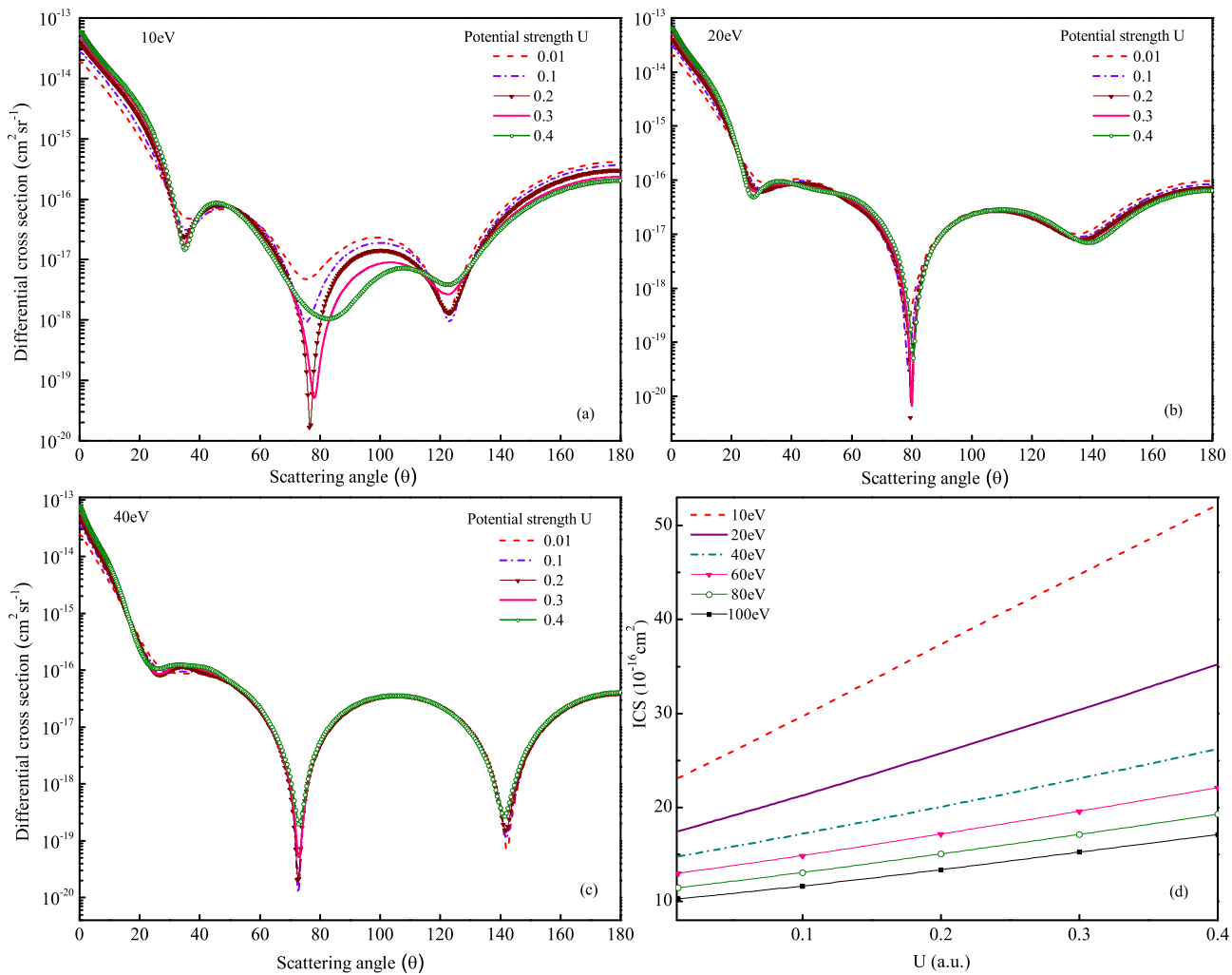}} 
\caption{(Color online) Demonstration of DCSs and ICSs using the DFCP method for the elastic scattering of an electron from a confined Ca atom 
as function of confining potential ($U$).}
\label{conf-ca}
\end{center}
\end{figure*}

\subsection{Elastic Scattering Cross-sections}

In this sub-section, we would like to demonstrate the effects of endohedral confinement of Ca on the electron elastic scattering cross-section. 
In order to assess the accuracy of our atomic structure calculations, we have calculated first DCSs for the electron and free Ca atom elastic 
scattering at the incident electron energies of 10, 20, 40, 60 and 100 eV and compare these calculations with the measurements reported by 
Milisavlijevi\'c et al \cite{Ca-expt} as well as those obtained using the optical model potential approach by Hasan et al \cite{hasan}. 
As mentioned in Sec. \ref{Model-Potential-theory}, we perform calculations using the charge densities from the DF and RCCSD methods.

In Fig. \ref{free-ca}, we plot our DCS results for the isolated Ca atom along with the other reported values in Refs. \cite{Ca-expt}
and \cite{hasan}. Calculations of Hasan et al \cite{hasan} are equivalent to our DFCP results and therefore, the two sets of results 
overlap on each other. At 10 eV, the RCCSD results explain the measurements better in comparison to the DFCP calculations in the scattering 
angle range 70$^o$ - 110$^o$. Also, the RCC results agree better with experimental data at 20 eV as well in the range 110$^o$ - 130$^o$. This 
indicates that the wave functions obtained using the RCCSD method give more reliable results at lower energies. At 20 eV, the DCS 
results are not reported by Hasan et al \cite{hasan} and hence, they are not shown in Fig. \ref{free-ca}(b). In general, our RCC and DFCP 
results are in good agreement with the measurements up to 70$^o$ electron scattering angle. Between 70$^o$ - 120$^o$ our results are slightly 
deviated in magnitude but agree in shape with the experimental values \cite{Ca-expt}. Beyond 120$^0$ our all the DCS, except at 60 eV, lie
within the experimental error bars. Further, as expected the correlation effects play important role in the forward scattering region and at low 
projectile electron energy. Therefore, as the energy increases the difference between the DF and DFCP calculations diminish. We also observed that 
the RCC results are slightly higher in the forward scattering range. This can be due to slightly higher value of $\alpha_0$ obtained by the 
RCCSD method in comparison to the DF method. In Fig. \ref{free-ca}(f), we have compared ICSs from our calculations and the experimental values. 
The DFCP results are in better agreement with the experimental data while cross-sections from the RCCSD method are slightly higher than 
the values obtained using the DFCP method. This is again due to higher values of DCSs from the RCCSD method at small angles. Agreement 
of the DFCP results with the experimental values could be coincidental, but we anticipate that inclusion of contributions from the higher level 
excitations in our RCC theory could improve the RCCSD results. Nevertheless, we find an overall agreement of the present calculations for 
electron and free Ca atom elastic scattering with the previous experimental \cite{Ca-expt} and other theoretical calculations \cite{hasan}.

Figure \ref{conf-ca} displays behavior of DCS as a function of confining potential $U$ in the electron energy range 10-100 eV. Although we 
have performed DCS calculations using the DF, DFCP and RCCSD methods, we present here only the DFCP calculations. We have used our calculated 
$\alpha_0$ values of the confined Ca atom to obtain polarization potentials at different $U$. 
It can be seen from the figure that the magnitude of DCS increases with increasing potential depth in the forward scattering direction. This 
reflects the fact that dipole polarizability of the target Ca atom increases with increasing well depth. Also, this could be due to rise 
in the charge density, as shown in Fig.~\ref{wavefunctions}, on increasing depth of the confining potential. The shape of the DCS curves at
various values of $U$ differ significantly at low incident electron energy. This difference diminishes with the increasing energy of 
the projectile electron. At 10 eV the first two dips near 35$^o$ and 80$^o$ become more pronounced and deeper with increasing $U$.   
In Fig-\ref{conf-ca} (d), the ICS are shown as function of well depth at various incident electron energy. As can be seen from the figure, 
there is a consistent increase in the cross-sections with increasing attractive potential. At low electron energy, the rise in ICS is more 
with increasing value of $U$ and the cross-section increases by 131\% at 10 eV energy for a change in the value of $U$ from 0.01 to 0.4 a.u..
However, this rise in ICS becomes slower for projectile electrons with more energy and for 100 eV ICS increases by 69\% only. Thus, the slope 
of these cross-section curves decreases with increasing electron energy. It shows that the influence of the presence of the endohedral 
environment on the cross-section is more significant for slower projectile.
 
\section{Conclusions}\label{conclusions}

We have carried out a detailed study of atomic structure and dynamics of endohedral confinement of Ca atom in C$_{60}$ cage using 
relativistic many-body methods. Since no experimental data is available for atomic properties of the above encapsulated atom, we have evaluated
self-consistent energy, correlation energy, ionization potential and dipole polarizability of this atom at different levels of 
approximations in the many-body methods. We found that deepening of the attractive potential well results into more tightening of 
the electrons to the nucleus and hence, increase in the ionization potentials. We also compared our single particle orbital energies 
for different values of potential depth with the corresponding results reported using numerical calculations and found very good agreement 
between them. Our ionization potential and dipole polarizability results for the free Ca atom agree well with the measurements. We also 
compared our self-consistent and correlation energies with the reported non-relativistic calculations and observe that the 
relativistic effects influence these results significantly.
  
The behavior of probability density and electronic charge density at various potential depths of the confined Ca atom obtained  
using the mean-field calculations are analyzed and it shows that these quantities show spectacular changes with strengthening of the 
confining potential. We found the probability density of the valence 4$s$ shell spreads over to a broader radial distance with increase 
in $U$ in contrast to the inner shells for which it remains unaffected in the presence of the potential well. Utilizing the 
electronic charge densities calculated by the Dirac-Fock method and relativistic coupled-cluster theory, the differential 
and integrated cross-sections of elastic scattering of an electron from the free and confined Ca atoms in the incident electron energy 
range from 10 to 100 eV are investigated. We found our results using the core-polarization potential with the Dirac-Fock wave function 
and wave functions from the relativistic coupled-cluster theory are in good agreement with the measurements and previous calculations
for the free Ca atom. The differential cross-sections are found to be slightly better in the relativistic coupled-cluster theory at the
incident electron energies 10 and 20 eV. This shows the effectiveness of the all-order perturbative coupled-cluster theory in the low energy regime.
It also emphasizes the crucial role played by the many-body methods for accurate determination of the atomic wave functions in explaining 
the experimental data. Further, variations in the differential and integrated cross-sections with increasing potential strength are 
shown. It shows differential cross-sections change dramatically; particularly at low projectile energies.

\section*{Acknowledgement}
SB is thankful to the Ministry of Human Resources and Development (MHRD), Govt. of India for fellowship. 
RS and LS are grateful to the SERB-DST and CSIR, New Delhi, Govt. of India for the supporting research grants. 
B. K. S. acknowledges use of Vikram-100 HPC cluster of Physical Research Laboratory (PRL), Ahmedabad, India for the computations.

\end{document}